\documentclass{ws-mpla}
\usepackage{psfig}
\begin{document}

\markboth{R Tharanath, Nijo Varghese and V C Kuriakose}
{ Phase transition, Quasinormal modes and Hawking radiation of
 Schwarzschild black hole in Quintessence field}

\catchline{}{}{}{}{}

\title{ Phase transition, Quasinormal modes and Hawking radiation of
 Schwarzschild black hole in Quintessence field}

\author{\footnotesize  R. Tharanath\footnote{tharanath.r@gmail.com},  Nijo Varghese\footnote{nijovarghesen@gmail.com}  and  V C Kuriakose\footnote{vck@cusat.ac.in}}

\address{Department of Physics, Cochin University of Science and
Technology, Kochi 682022, India\\
}

\maketitle

\pub{Received (Day Month Year)}{Revised (Day Month Year)}

\begin{abstract}

Black hole thermodynamic stability can be determined by studying the nature of 
heat capacity of the system. For Schwarzschild black hole 
the heat capacity is negative, but in the quintessence field this system shows a second order 
phase transition, implying the existence of a stable phase. We further discuss the equation 
of state of the present system. While analyzing
the quasinormal modes, we find that the massive scalar quasinormal mode  frequencies in the
complex $\omega$ plane shows a dramatic change when we plot it as a progressive function of quintessence
state parameter.
We also find the Hawking temperature of the system via the method of tunneling.
\end{abstract}
\ccode{PACS Nos.: 04.70.Dy, 04.70.-s }

\section{Introduction}\label{intro}

The thermodynamic properties of black holes have received considerable attraction in recent times,
as it is hoped that these studies can establish a connection among thermodynamics, gravitation and quantum statistical mechanics and 
eventually leading to quantum gravity.
Since the seminal works of Hawking\cite{hawking} and Bekenstein\cite{bekenstein,bekenstein2}, it is understood that black holes behave as 
thermodynamic objects, with characteristic temperature and entropy. Hawking radiation has not been yet directly observed but
the thermodynamic properties are thoroughly understood. 
  The realization that black hole laws are thermodynamic in nature
implies that there should be an underlying statistical description of them in terms of some microscopic states. 
Black hole thermodynamics is now widely studied. It is well known that, for Schwarzschild black hole
the heat capacity is negative and it is thermodynamically unstable.

    Accelerating expansion of the universe is a most recent fascinating result of observational cosmology. 
To explain the accelerated expansion of the universe, it is proposed that the universe is regarded as being dominated
by an exotic scalar field with a large negative pressure called ``dark energy'' which constitutes about 
70 percent of the total energy of the universe. There are several candidates for dark energy. ``Quintessence''\cite{Caldwell,Freese} is one among them.
It is characterized by a parameter $\epsilon$, the ratio
of the pressure to energy density of the dark energy, and the value of $\epsilon$ falls in the range $-1\leq\epsilon\leq-\frac{1}{3}$. 
  In our previous study\cite{tharanath} of 
Schwarzschild black hole surrounded by quintessence, we observe a second order thermodynamic phase transition for the black hole,
thus it possesses a positive heat capacity regime and thermodynamic stability.

   The study of quasinormal modes gained great attention since the existence of QNMs was first
pointed out by Vishveshwara\cite{vis} in the calculation of the scattering of gravitational
waves by a black hole. They are considered to be the characteristic sound of the black holes.
The quasi normal modes of different black holes surrounded
by quintessence have been studied earlier\cite{Songhai,Nijo}. 
Connection between black hole quasinormal modes and their phase transition was studied in\cite{jing}
and later it was found that the relation is not so trivial\cite{berti}. Here we are investigating the massive scalar QNMs
of the Schwarzschild-Quintessence black hole.

It will be very interesting to study the thermal emission from the Schwarzschild-Quintessence black hole.
Several derivations of Hawking radiation exist in literature. Parikh and Wilczek\cite{parikh} put forward a semi-classical
quantum tunneling model that implemented Hawking radiation as a tunneling process. More specifically they considered 
the effects of a positive energy matter shell propagating outward through the horizon of the Schwarzschild and Reissener-Nordstrom
black holes. The back reaction\cite{banerji2,majhi2} and noncommutative effects\cite{banerji3} have also been discussed by tunneling 
mechanicm. We find the Blotzman factor via the method of tunneling and study its variation with respect
to the quintessence state parameter.

The paper is organized as follows. In section \ref{ther} we discuss the second order phase transition and equation of state of the black hole.  
In section \ref{qnm} we  
calculate the QNMs of a massive scalar field and we observe a connection between QNMs and phase transition and 
in section \ref{haw} we calculate the Hawking radiation through tunneling mechanism. This paper ends with conclusion 
in section \ref{con}.

\section{Thermodynamics}\label{ther}
\subsection{Second order phase transition}\label{pt}

Phase transition is an important phenomenon in thermodynamics, so it is natural to probe the same in
black hole thermodynamics. The work of Hawking and Page\cite{hawkingpage} proved that there is a phase transition
between thermal AdS state and AdS black hole in 4 dimensions as the temperature changes. And later the black hole phase 
transition has been extended and indicates that there exist different phase transitions under various circumstances
\cite{davies1,chamblin,caldarelli,carter,dey,sheykhi,fernando,majhi}. 
 The phase transition is always identified with the sign change of heat capacity. Davies\cite{davies} argued that the point at which 
the specific heat travels from positive to negative values through an infinite discontinuity marks a 
phase transition.

The discovery of thermal emission of elementary particles by
Schwarzschild black holes has initiated deeper investigations of thermodynamic properties of stationary, rotating and
charged black holes. Those investigations studied stable equilibrium of non-rotating black holes with a 
thermal radiation bath\cite{hut,gibbons,hawking2}, the
fluctuation-dissipation theorem in irreversible thermodynamics\cite{candelas}, the black-hole version
for the third law of thermodynamics and specific heats of black holes in thermal equilibrium\cite{davies,hut}. 
It has been found that Black hole thermodynamics differs from the normal theory of 
thermodynamics in a number of ways: apart from the unsolved problem of a proper definition of stable equilibrium for Kerr black holes, 
Hawking\cite{hawking2} has shown that black holes cannot be
described by means of a canonical ensemble (this is closely related to the fact that the black-hole entropy is a 
global property, since it cannot be divided up into a number of weakly interacting parts).

\begin{figure}
 \centering
\includegraphics[width=0.60\columnwidth]{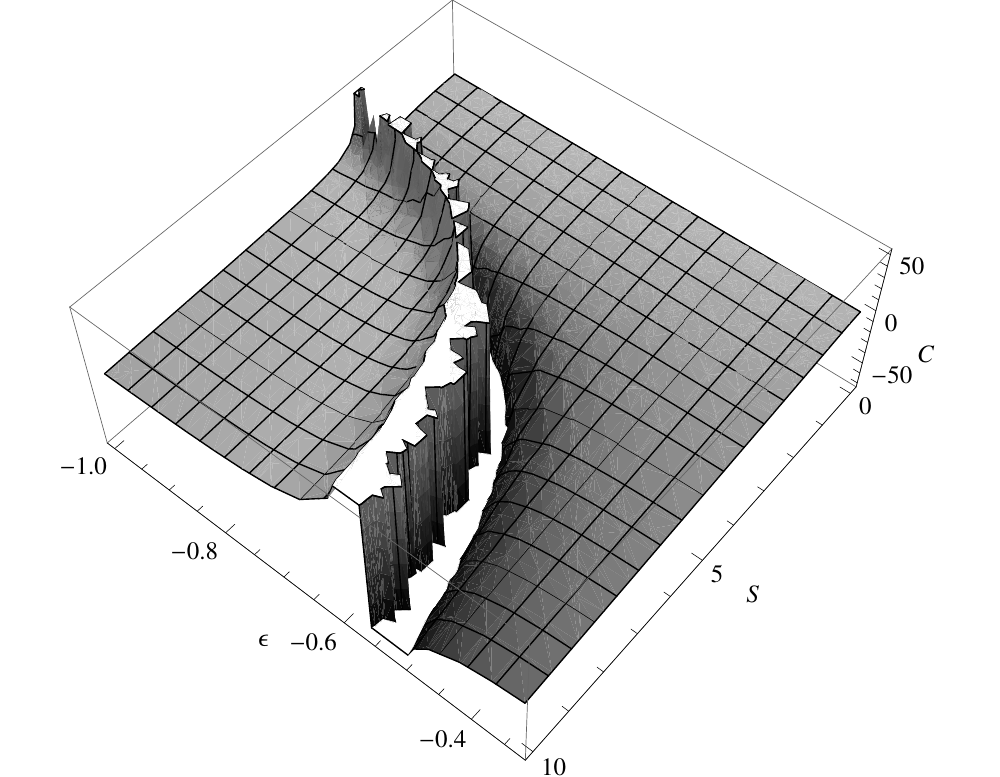}
\caption{Variation of heat capacity with entropy and with quintessence state parameter $\epsilon$ keeping $a=0.1$. }
\label{heat capacity}
\end{figure}

We are considering the Schwarzschild black hole surrounded by quintessence.
In the present study, the black hole is regarded as
a thermal system and it is then natural to apply the laws of thermodynamics. However,
a crucial difference from other thermal systems is that it is a gravitational object whose
entropy is identified with the area of the black hole(here we are using $c=G=\hbar=1$).
The metric of a Schwarzschild black hole surrounded by quintessence\cite{kiselev} is given by,
\begin{equation}\label{metric}
  ds^{2}=f(r)dt^{2}-\frac{1}{f(r)} dr^{2}-r^{2}(d\theta^{2}+\sin\theta^{2}d\phi^{2}),
\end{equation}
where
 \begin{equation}\label{f(r)}
  f(r)=1-\frac{2M}{r}-\frac{a}{r^{3\epsilon+1}}.
\end{equation}
Here M is the black hole mass and $a$ is the normalization factor, which is positive, depending on the energy density
 of quintessence. Quintessence is a scalar field whose equation of state parameter $\epsilon$ is defined as the ratio of its pressure $p$ and its
energy density $\rho$, which is given by a kinetic term and a potential term as\cite{Caldwell}, $\epsilon\equiv\frac{p}{\rho}=\frac{\frac{1}{2}\dot Q^{2}-V(Q)}{\frac{1}{2}\dot Q^{2}+V(Q)}$.
Following Kiselev\cite{kiselev}, the energy density can be written as $\rho_{\epsilon}=-\frac{a}{2}\frac{3\epsilon}{r^{3(1+\epsilon)}}$.

            We can establish the relation between mass of a black hole and its horizon radius directly from  
(\ref{f(r)}) as, 
\begin{equation}\label{mr}
 M=\frac{r}{2}-\frac{a}{2 r^{3\epsilon}},
\end{equation}
and we know that entropy can be written as
\begin{equation}\label{sa}
 S=\frac{A}{4}= \pi r^{2},
\end{equation}
so that $r$ can be written in terms of $S$ as 
\begin{equation}\label{rs}
 r= \sqrt{\frac{S}{\pi}}.
\end{equation}
Let us rewrite (\ref{mr}) using (\ref{rs}) as
\begin{equation}\label{ms}
 M=\frac{1}{2}\left[\sqrt{\frac{S}{\pi}} - a(\frac{\pi}{S})^{\frac{3 \epsilon}{2}}\right].
\end{equation}

Now we can deduce the heat capacity from the above expression for mass in terms of entropy.
Heat capacity in terms of entropy and quintessence parameter is given by
\begin{equation}\label{cs}
 C=T \frac{\partial S}{\partial T}=- \frac{16 S^{3 \epsilon +5}+ 96 a \epsilon \pi^{\frac{3\epsilon +1}{2}} S^{\frac{3 \epsilon +9}{2}}}{8 S^{3 \epsilon +4}+144 a\epsilon^{2} \pi^{\frac{3\epsilon +1}{2}} S^{3 \epsilon +2}+ 96 a \epsilon \pi^{\frac{3\epsilon +1}{2}} S^{\frac{3 \epsilon +7}{2}}}.
\end{equation}

In Fig.\ref{heat capacity} we have drawn the heat capacity as a three dimensional plot by introducing 
the quintessence state parameter along the third axis and it is clear that there is a second order phase transition.  From 
the plot we can find the critical point of phase transition for each value of 
quintessence state parameter. The quintessence effect in fact makes the thermodynamically unstable
 Schwarzschild system stable and changes the transition point with respect to the state parameter.
The 3 dimensional plot actually enables us to find
the dependence of heat capacity on the quintessence parameter. It is obvious that the infinite discontinuity 
has not been shown for all values of quintessence state parameters. From a certain value of quintessence parameter onwards the 
phase transition behaviour begins. We could see that for the Schwarzschild like case in the quintessence field, 
i.e., for $\epsilon= -\frac{1}{3}$, the heat capacity does not show any kind of phase transition. Thus it agrees with 
the existing results of Schwarzschild 
case. 
\subsection{Equation of state of the Black hole}\label{eq}

\begin{figure}
 \centering
\includegraphics[width=0.60\columnwidth]{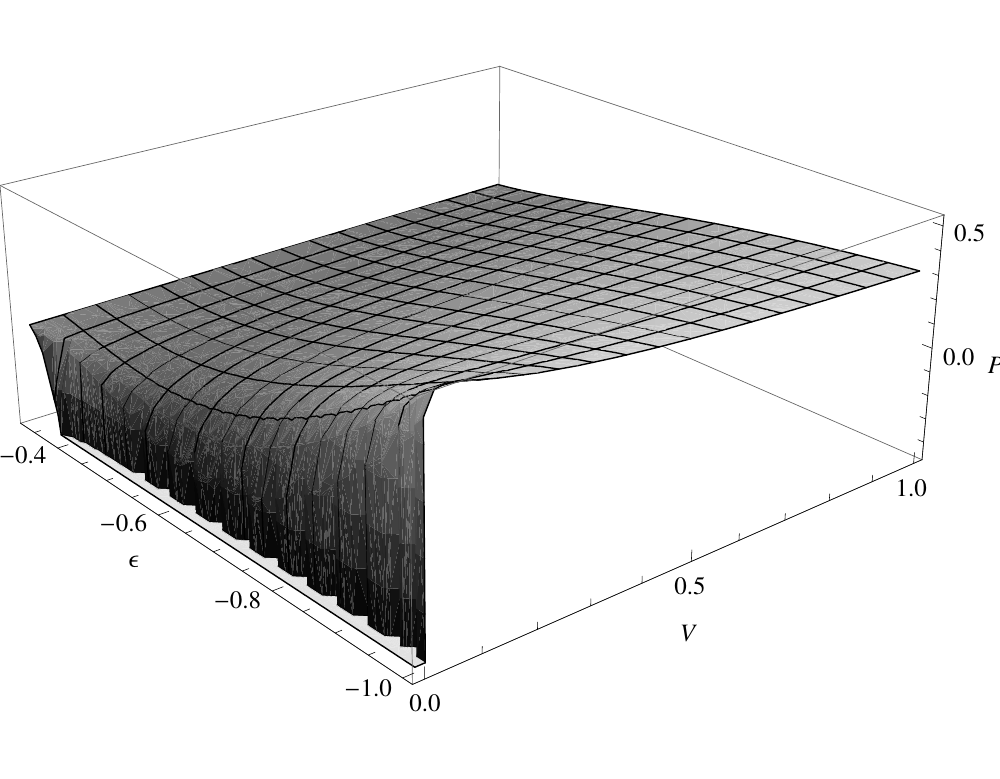}
\caption{P-V isotherms with quintessence state parameter as the third axis}
\label{pv}
\end{figure}

The cosmological constant related term in the metric, will act as a pressure term\cite{dolan1,dolan2}.
Thus we could write
\begin{equation}\label{p}
 P=-\frac{a}{8 \pi},
\end{equation}
and the mass of the black hole, $M$ is most naturally associated with the enthalpy $H$ of the black hole, hence
\begin{equation}
 H=E+PV.
\end{equation}

In black hole thermodynamics also, volume has been considered 
as a thermodynamic variable\cite{dolan1,cvetic}. So we find the volume of the 
black hole thermodynamically and find the equation of state.
The natural variables for enthalpy are entropy and pressure, so we could write $H$, in turn $M$, as a function 
of $S$ and $P$, 
\begin{equation}
 M=H(S,P).
\end{equation}
Now using (\ref{mr}) and (\ref{p}), enthalpy can be written as
\begin{equation}
 H(S,P)=\frac{1}{2}\left(\frac{S}{\pi}\right)^{\frac{1}{2}} 
\left[ 1+ \frac{8 \pi^{\frac{3 \epsilon+3}{2}}P}{S^{\frac{3\epsilon+1}{2}}}\right].
\end{equation}
We can find the volume of the Black hole using Legendre transformation,
\begin{equation}
 V=\left(\frac{\partial H}{\partial P}\right)_S=\frac{4 \pi}{r^{3 \epsilon}}.
\end{equation}
The equation of state of black hole can be written as,
\begin{equation}
 T=\frac{1}{4 \pi}\left[\left(\frac{V}{4 \pi}\right)^{\frac{1}{3 \epsilon}}-\frac{6 \epsilon P}{(4 \pi)^{\frac{2}{3 \epsilon}}}
V^{(1+ \frac{2}{3 \epsilon})}\right].
\end{equation}
We have plotted the P-V isotherms with the quintessence state parameter $\epsilon$ in Fig.\ref{pv}.

\section{Quasinormal modes and phase transition}\label{qnm}

The massive scalar field in a curved background is governed by the Klein-Gordon equation:
\begin{equation}\label{kg}
\Box{\Phi}-m^{2}\Phi=\frac{1}{\sqrt{-g}}(g^{\mu \nu}\sqrt{-g}\Phi_{,\mu})_{,\nu}-m^{2}\Phi=0,
\end{equation}
where $\Phi$ is the scalar field.

Using(\ref{metric}) in (\ref{kg}) and separating
angular and time variables, we obtain the radial equation:
\begin{equation}
\frac{d^{2}}{dr_{*}^{2}}+[\omega^{2}-V(r)]\Phi(r)=0,
\end{equation}
where,
\begin{equation}
V(r)=(1-\frac{2M}{r}-\frac{a}{r^{3\epsilon+1}})(\frac{l(l+1)}{r^{2}}+\frac{2M}{r^{3}}+\frac{a(3\epsilon+1)}{r^{3\epsilon+3}}+m^{2}),
\end{equation}
\begin{equation}
dr_{*}=\frac{dr}{1-\frac{2M}{r}-\frac{a}{r^{3\epsilon+1}}},
\end{equation}
and $l=0,1,2,3...$ parameterizes the field angular harmonic index.
The effective potential $V(r)$ approaches to a constant both at
the event horizon and at spatial infinity. It is clear that the
effective potential relates to the value of $r$, angular harmonic
index $l$, the state parameter $\epsilon$, the scalar field mass
$m$, the normalization factor $a$ and the mass of the black hole
$M$. However, in this paper, we only want to investigate the
relationship between the state parameter $\epsilon$ and the
scalar field mass $m$ with the quasinormal modes. Therefore, taking
$M=1$ and $a=0.1$, we compute the quasinormal frequencies
stipulated by the above potential using the third-order WKB method
developed by Schutz, Will and Iyer\cite{schutz,iyer1,iyer2}.

\begin{figure}
 \centering
\includegraphics[width=0.60\columnwidth]{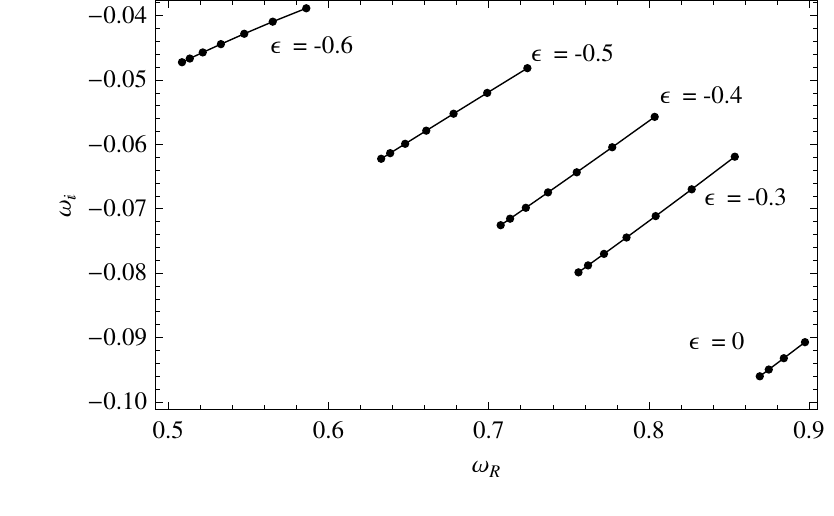}
\caption{Figure represents massive-scalar QNMs of Schwarzschild black hole surrounded by Quintessence,
with $l=4$, $n=0$, $a=0.1$ and we plot it for different values of mass( $m=0.1,0.2,0.3$ etc), ($\epsilon=0$ is the Schwarzschild case
with $a=0$).}
\label{qnmm}
\end{figure}

 Fig.\ref{qnmm} represents the quasinormal mode frequencies for different values of quintessence parameter, including the Schwarzschild case 
for which $a=0$. The values of QNMs for each quintessence state parameter and for different values of `$m$', is given in Tab.\ref{table1}.
 It is clear that the quintessence effect is to shift the frequencies away from the original Schwarzschild case. 
Now we are probing the QNM frequencies to get some notion about the phase transition,
which can be understood by plotting the complex frequencies with progressing values of quintessence parameter.
Tab.\ref{table2} gives the QNM frequencies for different values of quintessence parameter.

\begin{figure}
 \centering
\includegraphics[width=0.60\columnwidth]{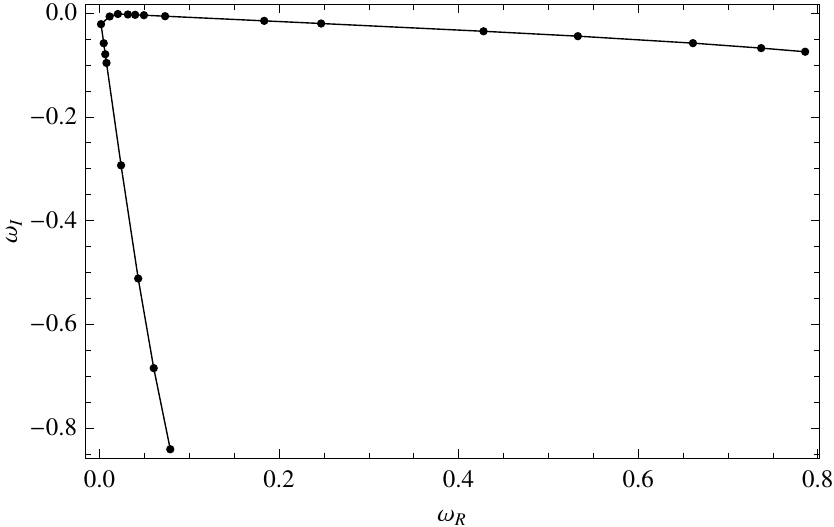}
\caption{Figure represents massive-scalar QNMs of Schwarzschild black hole surrounded by Quintessence,
with $l=4$, $n=0$, $a=0.1$,$m=0.4$ and we plot it for different values of quintessence state parameter $\epsilon$. Here $\epsilon=-0.33$ at the right extreme of the curve,
$\epsilon=-0.66$ at the turning point and it terminates at $\epsilon=-1$. }
\label{hc and qnm}
\end{figure}

Fig.\ref{hc and qnm} represents the QNM spectrum with respect to the varying quintessence state parameter keeping mass `$m$' fixed. We can
now see that the value of $\epsilon$ at which the heat capacity shows a phase transition(Fig.\ref{heat capacity}) coincides with the value of $\epsilon$ at which
the QNM spectrum showing a change in its slope. 
In the previous studies also, such a numerical coincidence has been found\cite{jing}. So we conclude that there may be a 
connection between thermodynamic and perturbative stabilities in the case of Schwarzschild black hole surrounded by quintessence.

The importance of this study lies on the fact that the phase transition is driven by the quintessence field. In the more generic case of quintessence field,
such as the Reissener-Nordstr\"{o}m-Quintessence black hole, the phase transition\cite{thomas} is mainly driven by the charge $Q$ and the quintessence state parameter $a$ has
got least significance. Of course it can be effective, when we use heavy quintessence field, but for any realistic case the quintessence densities
will be much lower than this. Thus, the similar study of connecting the phase transition and QNM spectra in the Reissener-Nordstr\"{o}m-Quintessence black hole will not make 
much difference from the work\cite{jing}. Whereas in the Schwarzschild-Quintessence black hole the system achieves the stable phase 
by the presence of quintessence only. So this study is quite important to check the influence of the quintessence field rather than that of charge. 

\begin{table}[h]
\tbl{Values of the quasinormal frequencies  for
low overtones($n=0$) in the Schwarzschild black hole($a=0$) and in the Schwarzschild black hole surrounded by
quintessence($a=0.1$) for fixed $l=4$.}
{\begin{tabular}{@{}cccccc@{}} \toprule
$a$ & $\epsilon$ & $\omega(m=0.1)$ &$\omega(m=0.2)$&$\omega(m=0.3)$ &$\omega(m=0.4)$
 \\

\colrule
0\hphantom{00} & \hphantom{0}0 & \hphantom{0}0.869210-0.096046i & 0.874830-0.094995i & 0.884224-0.093232i& 0.897437-0.090742i \\
0.1\hphantom{00} & \hphantom{0}-0.3 & \hphantom{0}0.756040-0.079890i & 0.762003-0.078819i&0.771979-0.077021i&0.786025-0.074475i \\
0.1\hphantom{00} & \hphantom{0}-0.4 & \hphantom{0}0.707477-0.072560i&0.713355-0.071551i&0.723185-0.069858i&0.737026-0.067465i \\
0.1\hphantom{00}&\hphantom{0}-0.5&\hphantom{0}0.632884-0.062241i&0.638489-0.061371i&0.647857-0.059917i&0.661032-0.057876i \\
0.1\hphantom{00}&\hphantom{0}-0.6&\hphantom{0}0.508557-0.047248i&0.513416-0.046673i&0.521512-0.045727i&0.532845-0.044432i\\
 \botrule
\end{tabular}\label{table1} }
\end{table}

\begin{table}[h]
\tbl{\label{def}Values of the quasinormal frequencies  for
low overtones($n=0$) in the Schwarzschild black hole surrounded by
quintessence($a=0.1$) for fixed $l=4$ and fixed mass ($m=0.4$).}
{\begin{tabular}{@{}cc@{}} \toprule
$\epsilon \hphantom{0000000000000000000000000000000}$ & $\omega_R + i \omega_I$
 \\

\colrule
 -0.3 \hphantom{00000000000000000000000000000000} &\hphantom{0}0.786025-0.074475i\\

-0.4 \hphantom{00000000000000000000000000000000} &\hphantom{0}0.737026-0.067465i\\

-0.5 \hphantom{00000000000000000000000000000000} &\hphantom{0}0.661032-0.057876i\\

-0.6 \hphantom{00000000000000000000000000000000} &\hphantom{0}0.532845-0.044432i\\

-0.7 \hphantom{00000000000000000000000000000000} &\hphantom{0}0.246904-0.020032i\\

-0.8 \hphantom{00000000000000000000000000000000} &\hphantom{0}0.043126-0.511825i\\

-0.9 \hphantom{00000000000000000000000000000000}&\hphantom{0}0.078842-0.841099i\\ \botrule
\end{tabular}\label{table2} }
\end{table}

\section{Hawking radiation via tunneling}\label{haw}
We present a short and direct derivation of Hawking radiation, considering it as a tunneling process based
on particles in a dynamical geometry for a Schwarzschild black hole surrounded by quintessence. 
To describe tunneling as an across horizon phenomena, it is necessary to choose coordinates which,
 unlike Schwarzschild coordinates, are not singular at the event horizon. 
Thus we rescale the time coordinate into Eddington-Finkelstein coordinates as
$t=T\pm r_{*}$, where the  $+$ and $-$ represent ingoing and outgoing particles respectively\cite{zhai,saleh}.
The tortoise coordinate $r_{*}$ is defined as, 
\begin{equation}
 \frac{dr_{*}}{dr}=f(r)^{-1}.
\end{equation}
In the following we study the outgoing particle only which is radiated from the black hole.
The background metric thus can be transformed to
\begin{equation}
 ds^{2}=-f(r) dT^{2} + 2dTdr+ r^{2} (d\theta^2+ sin^{2} \theta d\phi^2).
\end{equation}

The apparent horizon of the metric is given by the equation 
\begin{equation}\label{fm}
 f(r)=1-\frac{2M}{r}-\frac{a}{r^{3\epsilon+1}}=0
\end{equation}
In the absence of quintessence $(a=0)$, this equation arrives at the solution  $r=2M$; now we consider the quintessence field strength 
as small and thus we could treat the whole quintessence field as a perturbation to the original background
metric. Eventually the new horizon radius will be slightly modified from the original horizon radius as
\begin{equation}
 R=r+\delta.
\end{equation}
Substituting this in (\ref{fm}), 
we obtain
\begin{equation}
 1-\frac{2M}{r}\left(1-\frac{\delta}{r}\right)-\frac{a}{r^{(3\epsilon+1)}}\left(1-(1+3\epsilon)\frac{\delta}{r}\right)=0,
\end{equation}
which gives us
\begin{equation}
 \delta \backsimeq \frac{a}{r^{3 \epsilon}},
\end{equation}
in the first approximation.

The radial null geodesic is given by 
\begin{equation}
\dot{r}=\frac{dr}{dT}=\frac{1}{2}\left(1-\frac{2M}{r}-\frac{a}{r^{(3\epsilon+1)}}\right).
\end{equation}

When a particle of energy $E$ is radiated away from the black hole, $\dot{r}$ becomes
\begin{equation}\label{null}
 \dot{r}=\frac{1}{2}\left(1-\frac{2(M-E)}{r}-\frac{a}{r^{(3\epsilon+1)}}\right).
\end{equation}

The imaginary part of the action is 
\begin{equation}
 Im \mathcal{S} = Im \int P_r dr =Im \int \int dP_r dr = Im \int \int \frac{dH}{\dot{r}}dr.
\end{equation}

where we have used the Hamilton's equation $\frac{dH}{dp_r}=\dot{r}$ and \\
$H=M-E'\Rightarrow dH=-dE'$. Thus, the imaginary 
part of action takes the form
\begin{equation}
 Im \mathcal{S} = Im \int_{M}^{M-E}\int \frac{2 dr}{1-\frac{2M}{r}-\frac{a}{r^{3\epsilon+1}}}(-dE') ,
\end{equation}

\begin{equation}
 =Im \int \frac{2 E r dr}{(r-R)}.
\end{equation}

We use the method of tunneling to evaluate the integral over $r$ and obtain
\begin{equation}
 Im \mathcal{S} = (4 \pi R) E.
\end{equation}

Now using the WKB approximation, the rate of radiation is expressed as 
\begin{equation}
 \Gamma \varpropto e^{-2 Im \mathcal{S}}=e^{- \beta E}.
\end{equation}

where $\beta$ is, 
\begin{equation}
 \beta= \frac{1}{T}=8 \pi \left(r+\frac{a}{r^{3\epsilon}}\right).
\end{equation}

\begin{figure}
 \centering
\includegraphics[width=0.60\columnwidth]{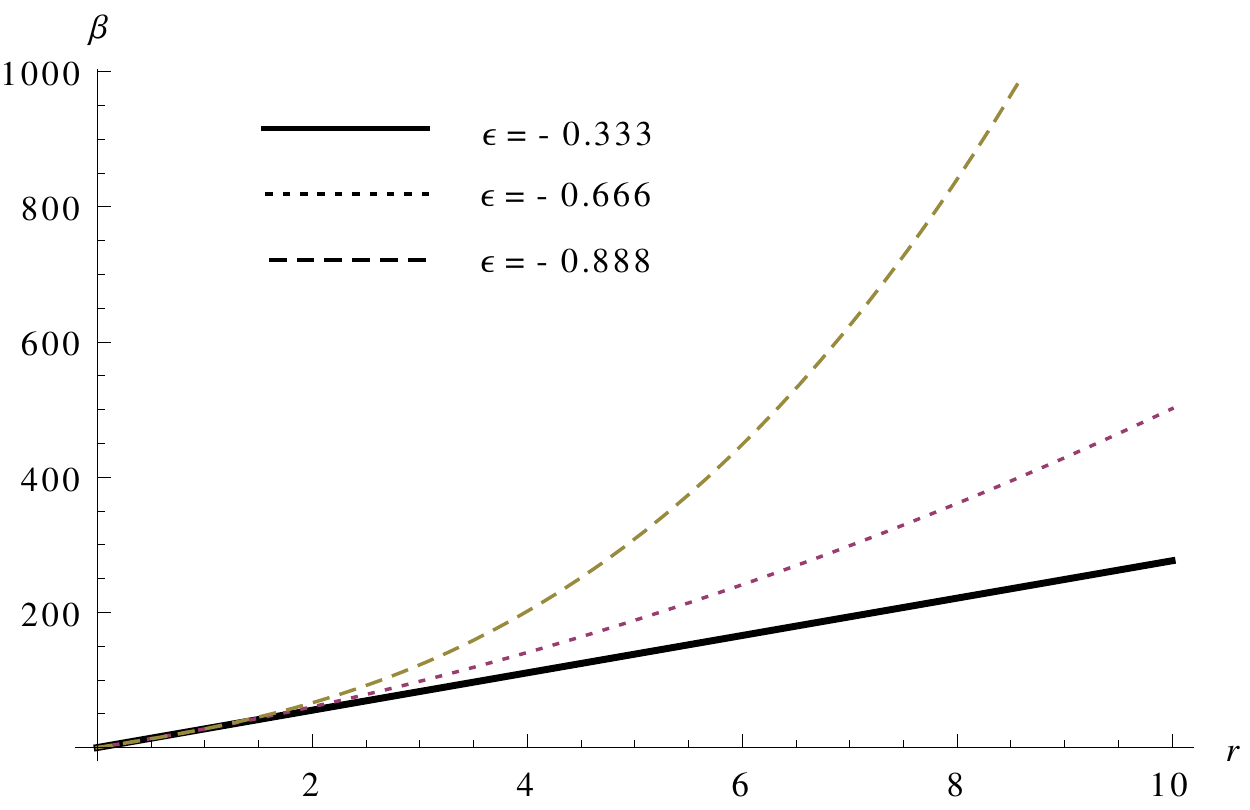}
\caption{Figure represents the variation of Boltzman factor with radius for different $\epsilon$}
\label{br}
\end{figure}

Fig.\ref{br} represents the variation of $\beta$ with respect to $r$. We can see that for different 
quintessence parameter values, $\beta$ diverges as $r$ increases.

\begin{figure}
 \centering
\includegraphics[width=0.60\columnwidth]{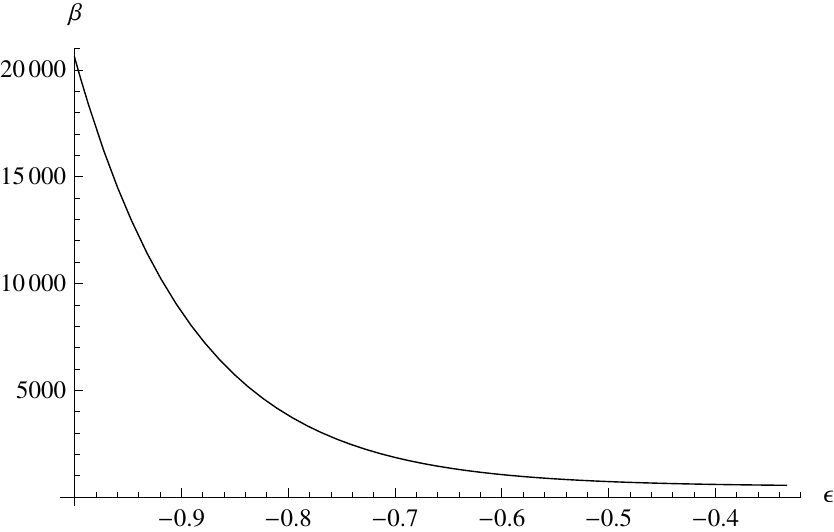}
\caption{Figure represents the variation of Boltzman factor with $\epsilon$ for different horizon radius}.
\label{hre}
\end{figure}

But the variation of $\beta$ with respect to the quintessence parameter $\epsilon$ is plotted in Fig.\ref{hre},
in which $\beta$ increases sharply below a particular value of $\epsilon$. This can be read along with the 
phase transition behaviour, which has been obtained in both the thermodynamic
 and perturbative approaches of sections \ref{pt} and \ref{qnm}.

Now we are going to find the change in entropy of the Black hole after emitting a particle out. 
For that we need to take the energy conservation into account. 
Then the radial null geodesic after emitting a particle of energy $E$ is given by (\ref{null}).

The imaginary part of the action of the massive particle is\cite{jiang}
\begin{equation}\label{ims}
 Im \mathcal{S}= Im \int_{t_i}^{t_f} L dt=Im \int_{r_{ie}}^{r_{fe}} (P_{r} \dot{r}) \frac{dr}{\dot{r}}
=Im \int_{r_{ie}}^{r_{fe}} \left[\int_{0}^{P_r} \dot{r} dP'_{r}\right] \frac{dr}{\dot{r}}.
\end{equation}

where $r_{ie}$ and $r_{fe}$ represent the localization of the event horizon before and after the emission of a particle
with energy $E$. $\dot{r}$ is given from the Hamilton's canonical equation of motion,

\begin{equation}\label{d}
 \dot{r}=\frac{dH}{dP_r}|_{r},~~~~~~~~~~~~~~~ dH|_{r}=d(M-E).
\end{equation}

Now substituting (\ref{null}) and (\ref{d}) in (\ref{ims}) we find,
 \begin{equation}
  Im \mathcal{S}=Im \int_{r_{ie}}^{r_{fe}} \int_{M}^{M-E}
\frac{ 2 \left[ d(M-E') \right] dr}{\left(1-\frac{2(M-E')}{r}-\frac{a}{r^{(3\epsilon+1)}}\right)},
 \end{equation}

which can be written as
\begin{equation}
 Im \mathcal{S}=Im \int_{r_{ie}}^{r_{fe}} \int_{M}^{M-E} \frac{2 r dr d(M-E')}{(r-R)},
\end{equation}

where $R=2(M-E)+\delta$. 

Now the second integral can be deformed as a contour, so as to ensure that positive energy solutions
decay in time. That is, we are taking the contour in the lower $E$ plane.

Using the method of Parikh and Wilczek\cite{parikh} we obtain
\begin{equation}
 Im \mathcal{S}= -Im \int_{r_{ie}}^{r_{fe}} r dr (\pi i)
= \frac{\pi}{2} (r_{ie}^2-r_{fe}^2).
\end{equation}

Using WKB approximation, we can get the tunneling rate of radiation as
\begin{equation}
 \Gamma \propto e^{-2Im\mathcal{S}}= e^{\pi(r_{ie}^2-r_{fe}^2)}=e^{\Delta S_{BH}},
\end{equation}
where $\Delta S_{BH}$ denotes the change in the Bekenstein-Hawking entropy 
at the event horizon before and after the particle tunnels out. It is obvious that the energy carried away by the tunneled particle will 
change the energy of the black hole and thus the entropy of the black hole should be decreased. In the perspective of area theorem,
the tunneling of particle results in decrease in the area as a few number of area quanta. The change in entropy found here 
can be quantized and in the semi-classical approach we can see that tunneling phenomenon and area quantization
give the same results.

The calculation of Hawking temperature via the tunneling method is also described in a more general way by Banerji et al\cite{banerji}, in which 
a general discussion of temperature for a general static, spherically symmetric black hole has been presented. The present 
expression can also be obtained from the general expression. 

\section{Summary and conclusion}\label{con}

In an earlier work\cite{tharanath}, we found that Schwarzschild black hole can have a stable phase when it is immersed in 
quintessence field. Here we analyze the second order thermodynamic phase transition in detail. We first plot the heat capacity in 3 dimensions taking the 
quintessence state parameter as one of the axes(Fig.\ref{heat capacity}). From which we could find the critical point changes as $\epsilon$ changes. It is evident from the plot that
for certain values of $\epsilon$(lower values of $\epsilon$, between $-\frac{1}{3}$ to $-\frac{2}{3}$), there is no phase transition. 
It is in general agreement with the result of Schwarzschild
case( i. e.,$\epsilon=-\frac{1}{3}$) that the system does not show any phase transition. 

Then we analyzed QNMs for the massive-scalar field of the same system(here we have used the same value of $a$, which we used to find its thermodynamic phase transition).
The complex frequency plot for different values of quintessence parameter
does not give any striking evidence of the phase transition we observed, but when we plot the imaginary frequencies as a
progressing function of quintessence parameter, we could see a turning point in the plot(Fig.\ref{hc and qnm}). The value of quintessence state
parameter $\epsilon$, at which the plot shows a change in slope coincides with the value of $\epsilon$ at which the heat capacity started showing phase transition(Fig.\ref{heat capacity}).
We have made a thorough investigation on the 
pressure and volume of the same system, and derived the equation of state.

We have also found 
the Hawking radiation via the method of tunneling for the same system. We have plotted the Boltzman factor as a function 
of both horizon radius and quintessence state parameter. The plot of $\beta$ verses $\epsilon$ also implies an indication of phase transition(Fig.\ref{hre}).

In summary, the present study shows that the value of quintessence state
parameter $\epsilon$ at which the heat capacity shows a phase transition coincides with the value of $\epsilon$ at which
the QNM spectrum showing a change in its slope. In the case of Hawking radiation, the plot of $\beta$ verses $\epsilon$ also shows 
a significant change at the same value of $\epsilon$. According to Berti\cite{berti}, the connection between QNMs and phase transition is not so trivial. 
But we could see a coincidence in the values of quintessence state parameter in thermodynamic phase transition, complex QNM spectrum and Hawking radiation.

\section{Acknowledgments}

TR wishes to thank UGC, New Delhi for financial support
under RFSMS scheme. NV Wishes to thank UGC, for the financial support under 
Kothari fellowship scheme. VCK is thankful to UGC, New Delhi for financial
support through a Major Research Project and wishes to acknowledge
Associateship of IUCAA, Pune, India.


\end{document}